# Scaling rules in the science system: influence of field-specific citation characteristics on the impact of research groups

**Anthony F. J. van Raan**
Center for Science and Technology Studies
Leiden University
Wassenaarseweg 52
P.O. Box 9555
2300 RB Leiden, Netherlands

***Abstract***
*We propose a representation of science as a citation-density landscape and investigate scaling rules with the field-specific citation density as a main topological property. We focus on the size-dependence of several main bibliometric indicators for a large set of research groups while distinguishing between top-performance and lower performance groups. We demonstrate that this representation of the science system is particularly effective to understand the role and the interdependencies of the different bibliometric indicators and related topological properties of the landscape.*

## Introduction

Science can be considered as a system of highly interconnected entities (e.g., individual publications, researchers, research groups, universities) that produce and transfer knowledge ('cognitive ecosystem'). Particularly important in such complex, large networked systems are the relations between large-scale attributes (in science for instance characteristics of fields) and local patterns (for instance the performance of individual groups). Most complex networks are the result of a growth process. This is certainly also the case for science with its increase of more than a million publications and twenty million citations each year. Topological properties of complex network systems are the fingerprints of the evolution of the system (Caldarelli, Erzan, & Vespignani 2004) though these properties themselves may remain remarkably constant.

There is a long history of the construction of bibliometric indicators (van Raan 2004) and there is much recent work on the use of publication and citation data in the study of author-, publication- and citation-networks in science (Albert & Barabási 2002; Dorogovtsev & Mendes 2002; Leicht, Clarkson, Shedden, & Newman 2007). But there is little work on the mutual coherence of bibliometric indicators and their statistical properties in the context of science as an interconnected system. Building on previous published work (van Raan 2006a,b) we continue in this paper our exploration of these interdependencies of the science system as a landscape characterized by field-specific citation densities.

In many networked systems a modular ordering or community structure is visible: the division of network nodes into groups within which the network connections are dense, but sparser between these groups (Klemm & Eguíluz 2002; Newman 2001, 2002, 2004; Newman & Girvan 2004; Ravasz & Barabási 2003). This is however not a recent discovery. In an early paper in Nature of this author (van Raan 1990) the structure of the science system was described as a natural phenomenon in which growth was found to be cluster-like. Statistical properties of co-citation clusters were analyzed and it was observed that the science system can be described in terms of fractal dimensions. The size distribution of the analyzed clusters is a snapshot of a dynamical process, reflecting the presence of established communities and the emergence of new ones. To our best knowledge, it was the first time that the idea of science as a cognitive ecosystem with research communities as 'species of scientists' was coined. Ten years later, we presented a model for the fractal differentiation of science (van Raan 2000) in which the power-law distribution of the total set of existing and emerging communities in the science system was explained on the basis of an ecological model of growth, ageing and creation. For early observations of power law characteristics of the science system we refer to Lotka (1926), Naranan (1971), Price (1976), and Haitun (1982), and for more recent work to Seglen (1992, 1994) and Redner (1998).

The ability to analyze statistical properties of groups within a network structure can provide new insights into our understanding of complex, network systems. Bibliometric analysis may provide new ways to quantitatively characterize this community structure in terms of further statistical properties of these communities and how these properties can be reconciled with the properties of the system as a whole. The bibliometric approach is particularly important because it provides the possibility to study 'science metrics' and in particular the modular structure of a real-world, complex network system with a known community structure in different hierarchies that are interrelated in a well-defined way. For instance, at the lowest level individual publications and their citation relations, next the individual researchers as authors and co-authors of publications, and at higher aggregation levels research groups, universities, and so on. Authors, groups, universities can be considered as 'modules' in a publication-based network as they all are in a bibliometric sense sets of related publications, be it of different size and different internal coherence.

All these entities are connected by two different types of interrelations: 'metabolic' links defined by citations, and semantic links defined by concept-similarities (van Raan & Noyons 2002; Menczer 2004). Both linkage types provide a metric, i.e., a measure for distances in the abstract space of the science system landscape. Particularly citation links enable us to study dynamical, time-dependent properties of the system. Thus, bibliometric data enable us to simultaneously capture topological properties (such as field-specific citation density) as well as the time-resolved dynamical processes (such as citations) taking place in the system, as was shown in our earlier work (van Raan 2000). Börner, Maru, & Goldstone (2004) recently published a model to describe the simultaneous growth and dynamic interactions of author- and publication-networks. Given the richness of available data and indicators, the bibliometric approach may also contribute to study universality of specific topological properties of complex systems, a leading question in network research (Caldarelli, Erzan, & Vespignani 2004).

In our previous paper (van Raan 2006b) we distinguished between top-performance and lower performance research groups in the analysis of statistical properties of bibliometric characteristics of research groups. The crucial finding was that particularly the lower performance groups have a size-dependent (size of a research group in terms of number of publications) *cumulative* advantage[1] advantage for receiving citations. Our goal in this paper is to investigate this scaling behavior further by focusing on the size-dependency of the main bibliometric indicators for different levels of *field-specific citation densities* while distinguishing between higher and lower performance groups.

The research group is the most important working floor entity in science. However, obtaining data at the research group level is by far a trivial matter. Data on the level of the individual scientists, institutions, and research fields are externally available (e.g., author names, addresses, journals, field classifications, etc.). But this is not the case at the level of research groups. The only possibility to study bibliometric characteristics of research groups with 'external data' would be to use the address information within the main organization, for instance 'Department of Biochemistry' of a specific university. However, the delineation of departments or university groups through externally available data such as the address information in international literature databases is very problematic (van Raan 2005). Furthermore, the external data has to be combined carefully with 'internally stored' data (such as personnel belonging to specific groups). These data are only available from the institutions that are the target of the analysis. As indicated above, the data used in this study are the results of evaluation studies and are therefore based on strict verification procedures in close collaboration with the evaluated groups.

The structure of this paper is as follows. In the second section we discuss the data material, the application of the method and the calculation of the indicators. We also introduce a representation of science as a 'citation density landscape'. This representation is particularly useful to understand the role and the interdependencies of the different bibliometric indicators. In the third section we present the results of our data analysis for 'external' (i.e., non self-) citations, and in the fourth section we discuss the main outcomes of this study in the framework of our landscape model.

## Data, Indicators, Citation-Density Landscape

The data material is based on a large set of publications (as far as published in journals covered by the Citation Index, 'CI publications'[2]) of all academic chemistry research in a country (Netherlands) for a 10-years period (1991-2000). This material is quite unique. To our knowledge no such compilations of very accurately verified publication sets on a large scale are used for statistical analysis of the characteristics of the indicators at the research group level. The (CI-) publications were collected as part of a large evaluation study conducted by the Association of Universities in the Netherlands. For a detailed discussion of the evaluation procedure and the results we refer to the evaluation report (VSNU 2002). In the framework of this evaluation study, we performed an extensive bibliometric analysis to support the evaluation work by an international peer committee (van Leeuwen, Visser, Moed, & Nederhof 2002). In total, the analysis involves 700 senior researchers and covers about 18,000 publications and 175,000 citations (excluding self-citations) of 157 chemistry groups at ten universities.

---

[1] With 'cumulative advantage' we mean that the dependent variable (for instance, number of citations of a group, ***C***) scales in a disproportional, non-linear way (in this case: power law) with the independent variable (for instance, in the present study the 'size' of a research group, in terms of number of publications, ***P***). Thus, larger groups (in terms of ***P***) do not just receive more citations (as can be expected), but they do so increasingly more 'advantageously': groups that are twice as large as other groups receive, for instance 2.4 times more citations. For a detailed discussion we refer to our previous paper (van Raan 2006b). For a general discussion of cumulative advantage in science we refer to Merton (1968, 1988) and Price (1976).

[2] Thomson Scientific, the former Institute for Scientific Information (ISI) in Philadelphia, is the producer and publisher of the Web of Science that covers the Science Citation Index (-extended), the Social Science Citation Index and the Arts & Humanities Citation Index. Throughout this paper we use the term 'CI' (Citation Index) for the above set of databases.

The indicators are calculated on the basis of a total time-period analysis. This means that publications are counted for the entire 10-year period (1991-2000) and citations are counted up to and including 2000 (e.g., for publications from 1991, citations are counted from 1991 to 2000, and for publications from 2000, citations are counted only in 2000). We applied the CWTS standard bibliometric indicators. Here only 'external' citations, i.e., citations corrected for self-citations[3], are taken into account. We present the standard bibliometric indicators with a short description in the text box here below. For a detailed discussion we refer to Van Raan (1996, 2004).

**Standard Bibliometric Indicators:**

- Number of publications ***P*** in CI-covered journals of a research group in the specified period;
- Number of citations ***C*** received by ***P*** during the specified period, without self-citations; including self-citations: ***Ci,*** i.e., number of self-citations ***Cs = Ci – C***, relative amount of self-citations ***Cs/Ci***;
- Average number of citations per publication, without self-citations (***CPP***);
- Percentage of publications not cited (in the specified period) ***Pnc***;
- Journal-based worldwide average impact as an international reference level for a research group (***JCS***, journal citation score, which is our journal impact indicator), without self-citations (on a world-wide scale!); in the case of more than one journal we use the average ***JCSm***; for the calculation of ***JCSm*** the same publication and citation counting procedure, time windows, and article types are used as in the case of ***CPP***;
- Field-based[4] worldwide average impact as an international reference level for a research group (***FCS***, field citation score), without self-citations (on a world-wide scale!); in the case of more than one field (as almost always) we use the average ***FCSm***; for the calculation of ***FCSm*** the same publication and citation counting procedure, time windows, and article types are used as in the case of ***CPP***; we refer in this article to the ***FCSm*** indicator as the 'field-specific citation density';
- Comparison of the ***CPP*** of a research group with the world-wide average based on ***JCSm*** as a standard, without self-citations, indicator ***CPP/JCSm***;
- Comparison of the ***CPP*** of a research group with the world-wide average based on ***FCSm*** as a standard, without self-citations, indicator ***CPP/FCSm***;
- Ratio ***JCSm/FCSm*** is the relative, field-normalized journal impact indicator.

In Table 1 we show as an example the results of our bibliometric analysis for the most important indicators for all 12 chemistry research groups of one of the ten universities (University A, groups A-01 to A-12). Table 1 shows that our indicator calculations allow a statistical analysis of these indicators. We regard the internationally standardized (field-normalized) impact indicator ***CPP/FCSm*** as our 'crown' indicator. This indicator enables us to observe whether the performance of a research group is significantly far below (indicator value < 0.5), below (0.5 - 0.8), around (0.8 - 1.2), above (1.2 – 1.5), or far above (>1.5) the international (western world dominated) impact standard of the field.

[3] A citation is a self-citation if any of the authors of the citing paper is also an author of the cited paper.
[4] We here use the definition of fields based on a classification of scientific journals into *categories* developed by Thomson Scientific/ISI. Although this classification is not perfect, it provides a clear and 'fixed' consistent field definition suitable for automated procedures within our data-system.

***Table 1****: Example of the results of the bibliometric analysis*

| *Research group* | *P* | *C* | *CPP* | *JCSm* | *FCSm* | *CPP/JCSm* | *CPP/FCSm* | *JCSm/FCSm* | *Cs/Ci* |
|---|---|---|---|---|---|---|---|---|---|
| | | | | | | | | | |
| A-01 | 92 | 554 | 6.02 | 5.76 | 4.33 | 1.05 | 1.39 | 1.33 | 0.23 |
| A-02 | 69 | 536 | 7.77 | 5.12 | 2.98 | 1.52 | 2.61 | 1.72 | 0.16 |
| A-03 | 129 | 3,780 | 29.30 | 17.20 | 11.86 | 1.70 | 2.47 | 1.45 | 0.16 |
| A-04 | 80 | 725 | 9.06 | 8.06 | 6.25 | 1.12 | 1.45 | 1.29 | 0.27 |
| A-05 | 188 | 1,488 | 7.91 | 8.76 | 5.31 | 0.90 | 1.49 | 1.65 | 0.30 |
| A-06 | 52 | 424 | 8.15 | 6.27 | 3.56 | 1.30 | 2.29 | 1.76 | 0.30 |
| A-07 | 52 | 362 | 6.96 | 4.51 | 5.01 | 1.54 | 1.39 | 0.90 | 0.16 |
| A-08 | 171 | 1,646 | 9.63 | 6.45 | 4.36 | 1.49 | 2.21 | 1.48 | 0.23 |
| A-09 | 132 | 2,581 | 19.55 | 15.22 | 11.71 | 1.28 | 1.67 | 1.30 | 0.25 |
| A-10 | 119 | 2,815 | 23.66 | 22.23 | 14.25 | 1.06 | 1.66 | 1.56 | 0.17 |
| A-11 | 141 | 1,630 | 11.56 | 17.83 | 12.30 | 0.65 | 0.94 | 1.45 | 0.29 |
| A-12 | 102 | 1,025 | 10.05 | 10.48 | 7.18 | 0.96 | 1.40 | 1.46 | 0.34 |

Particularly with a ***CPP/FCSm*** value above 1.5, groups can be considered as scientifically strong. A value above 2 indicates a very strong group and groups with values above 3 can generally be considered as excellent and comparable to top-groups at the best US universities (van Raan 2004). The ***CPP/FCSm*** indicator generally correlates well with the quality judgment of the peers (van Raan 2006a, b). Studies of large-scale evaluation procedures in which empirical material is available with data on both peer judgment as well as bibliometric indicators are quite rare. For notable exceptions, see Rinia, van Leeuwen, van Vuren, & van Raan (1998, 2001).

We observe in Table 1 large differences in the ***FCSm*** values for the various research groups. This clearly illustrates that research groups even within one discipline (in this case chemistry) may work in fields with a high or a low field citation density. Generally we find high field-specific citation densities in basic research and low field-specific citation densities in applied research. For instance, research group A-02 is active in an applied field, catalysis research, and this group is characterized by a low field-specific citation density (***FCSm*** = 2.98). Group A-10 focuses on medicine-related basic research on human proteins, this group has a typical high field-specific citation density: ***FCSm*** = 14.25. For the total set of research groups we find that the ***FCSm*** values may differ a factor of about 20, so more than an order of magnitude. Thus, these findings show that the idea of science as large collection of research groups positioned in a 'citation density landscape' makes sense.

In Fig. 1 we give a simple representation of the science system at the lowest aggregation level: a networked system of connected publications in a landscape with different citation densities, i.e., publications connected by citations and embedded in larger entities such as journals and fields. Schematically we illustrate the meaning of our ***CPP***, ***JCSm***, and ***FCSm*** indicators: citations per publication, citations to the set of publications in a journal, and citations to the entire set of publications that form together a specific field, respectively.

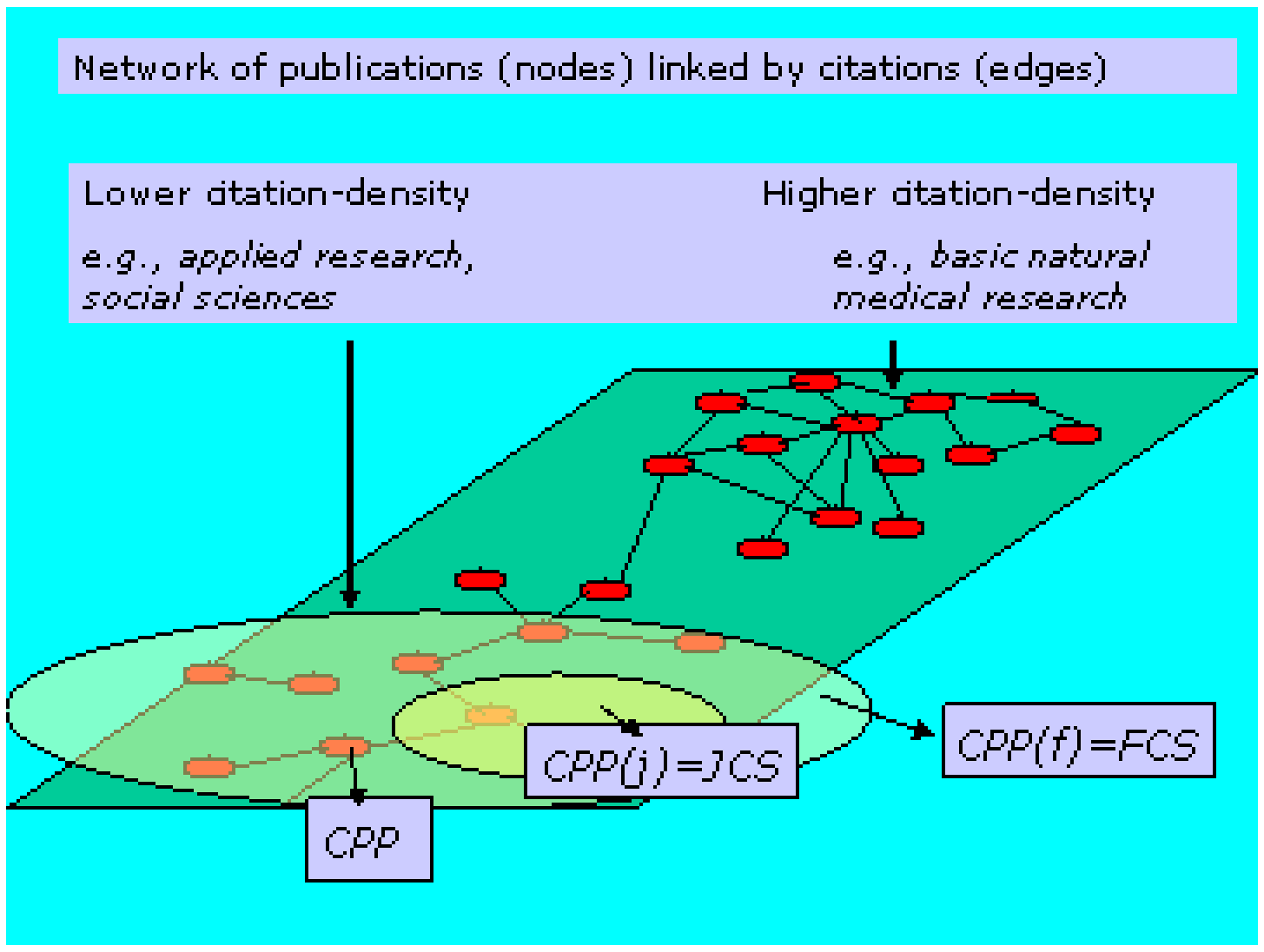

***Figure 1****:* *Science as a networked system of connected publications with a schematic illustration of the concept of citation-density and with of our* ***CPP****,* ***JCSm*** *and* ***FCSm*** *indicators.*

In this 'science system landscape' the field citation density[5] ***FCSm*** is the main topological parameter defining the *coarse-scale structure* of the landscape. The journal citation density ***JCSm*** is as it were a fine-tuning within this coarse scale structure and compared to the ***FCSm*** it is a more group-specific attribute. From this perspective we can visualize the position of a research group –formally as a set of publications within one or several fields- in the science landscape as presented in Fig. 2: groups in regions (i.e., fields of science) in the landscape with high and low citation densities in which the indicators can be illustrated relative to each other.

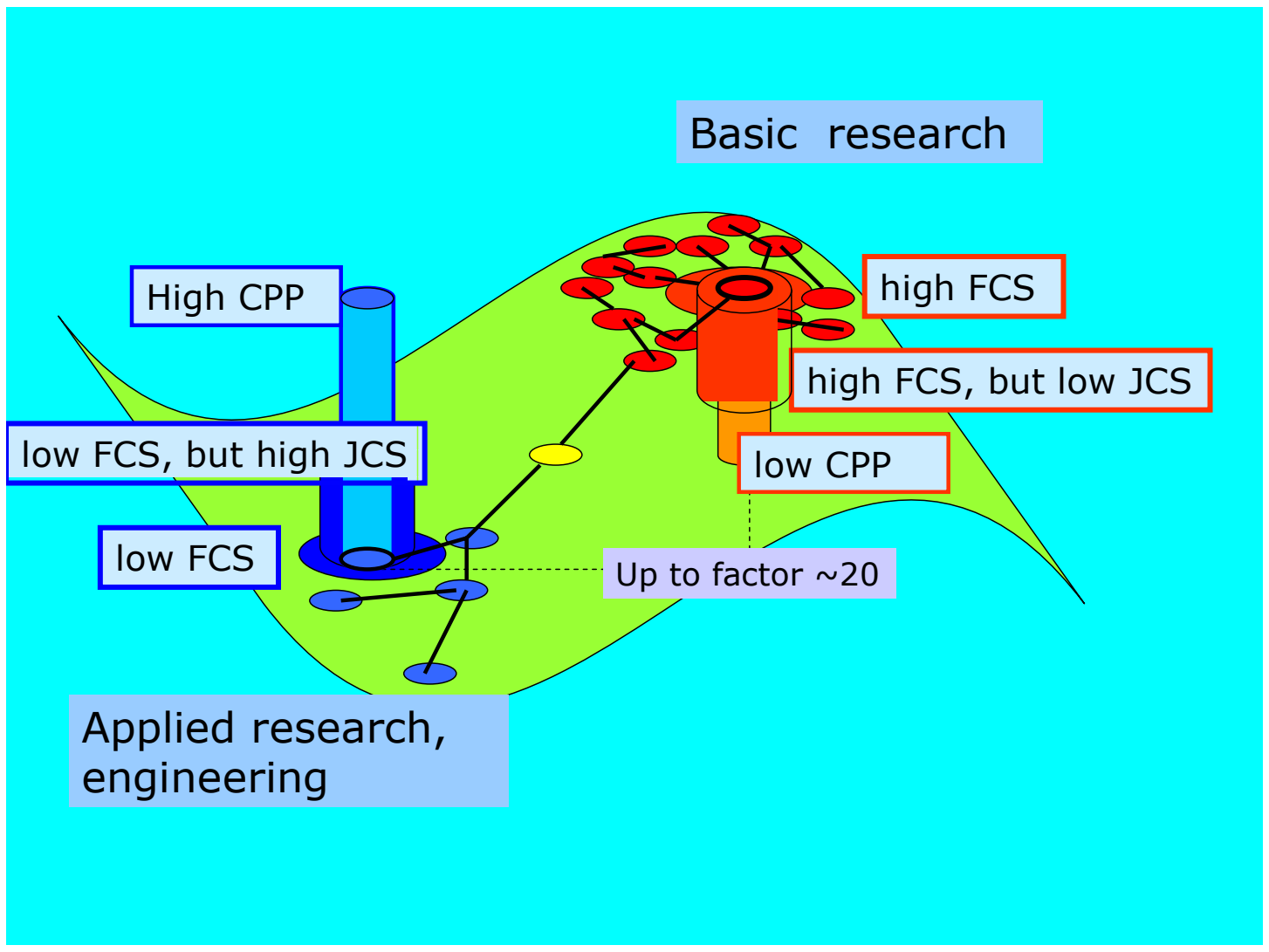

***Figure 2****:* *The science system from the perspective of field-specific citation-densities with two groups as examples of the* ***CPP****,* ***JCSm*** *and* ***FCSm*** *indicators.*

[5] In the remainder of this paper we will leave out 'specific' and use the simpler concept 'field citation density'.

In the lower part of the landscape we see a group with a low ***FCSm***, but publishing in the better journals of the field, which means that ***JCSm*** > ***FCSm***, and within these top-journals the group performs very well so that ***CPP*** > ***JCSm***. In Table 1 we see that research group A-02 is an example of this situation.

## Results and discussion

### *Influence of field-specific citation density and journal impact*

In Fig. 3 we present the correlation of the number of citations with size in terms of number of publications (***P***) for the groups with *high* and *low* field citation-densities, i.e., the top-25% and bottom-25%, respectively, of the ***FCSm*** distribution. The figure shows that there is a cumulative 'disadvantage' for the high field density groups (power law exponent $\alpha$ = 0.84) and a considerable cumulative advantage for the low field-specific citation density groups (power law exponent $\alpha$ = 1.41).

Thus, for larger sizes (high ***P***) the difference in number of citations between high and low field density groups will become smaller. This is clearly illustrated by Fig. 4 where we show the correlation between ***CPP*** and ***P*** for the high and the low field citation-density groups. We observe in Figs. 3 and 4 a convergence of the impact (***C***, as well as ***CPP***) of the low and the high field citation–density groups. More precisely, Fig. 5 shows a convergence of the *high* field citation-density *top*-performance groups with the *low* field citation-density *top*-performance groups around ***P ~*** 1,000 with ***CPP*** ~ 15, and a convergence of the *high* field citation-density *lower* performance groups with the *low* field citation-density *lower* performance groups also around ***P ~*** 1,000 with ***CPP*** ~ 10. Clearly, for very large sizes, groups cover a broad range of fields and the distinction between high and low field-citation density will lose its meaning as their average ***FCSm*** will tend to the same value.

We observe that the average number of citations in the *highest* 25% ('top') regions of the field citation-density landscape ($CPP_t$) relates to size approximately as

$$CPP_t \sim P^{-0.2},$$

whereas the average number of citations in the *lowest* 25% ('bottom') regions of the field density landscape ($CPP_b$) relates to size approximately as

$$CPP_b \sim P^{+0.4}$$

This is quite a remarkable difference, as from the above equations simply follows that -again in first approximation-

$$CPP_b \sim (CPP_t)^{-2}$$

This clearly illustrates the importance of field citation densities in the bibliometrically constructed science system. More specifically, these observations directly relate to a well-know question: does a larger number of publications lead to a 'dilution' of the average impact (citations) per publication? Our empirical results clearly provide an answer to this question. The answer is yes, for the *high* field citation-density groups, and no for the *low* field citation-density groups. But how does research performance relates to this phenomenon?

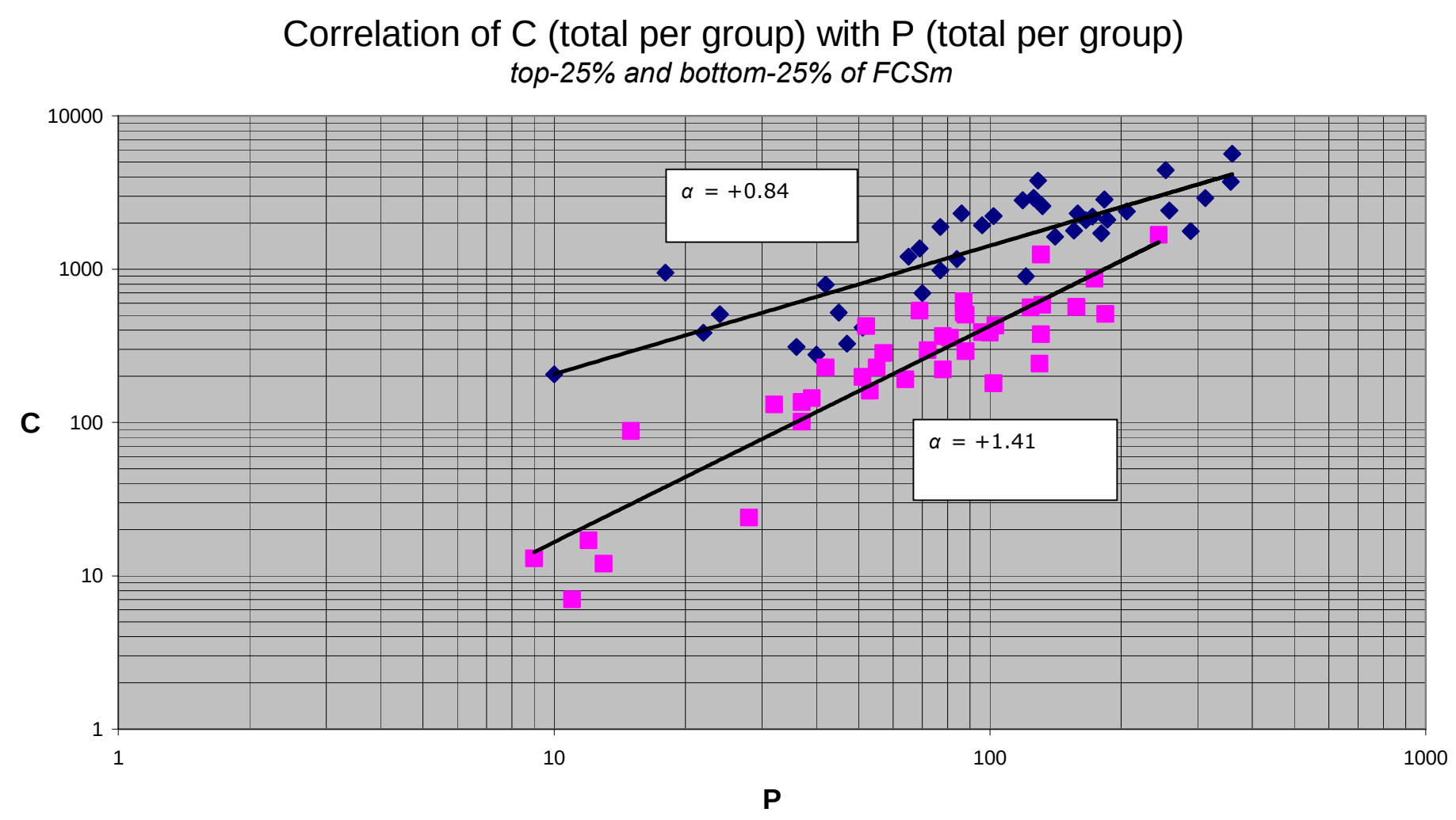

***Figure 3:*** *Correlation of the number of citations (**C**) with the number of publications (**P**), for groups in the top-25% (diamonds) and in the bottom-25% (squares) of the field citation-density (**FCSm**) distribution.*

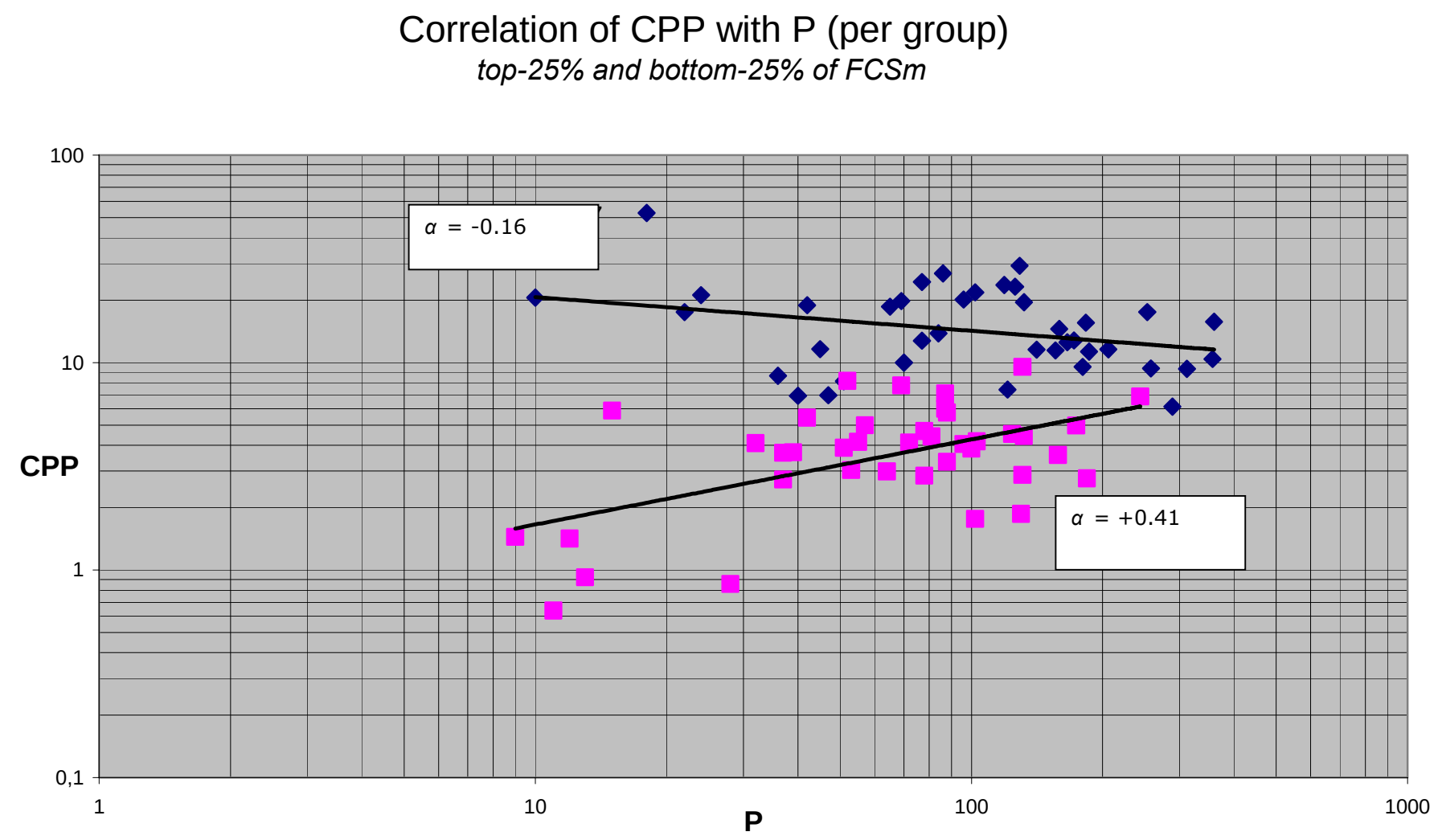

***Figure 4:*** *Correlation of citations-per-publication (**CPP**) with the number of publications (**P**) for groups in the top-25% (diamonds) and in the bottom-25% (squares) of the field citation-density (**FCSm**) distribution.*

In Fig. 5 we present the same data as in Fig. 4, but now we distinguish within the high field citation-density groups between high and low performance groups (i.e., the top-25% and the bottom-25% of the ***CPP/FCSm*** distribution, respectively), and the same for the low field citation-density groups. We clearly see the difference in ***CPP*** for the high and the low performance groups in case of high and low field citation-density. We again observe that only the low field citation-density groups benefit from a larger size, and this is the case for both the high and low performance groups. In these *low* field citation-density groups, the *high* performance groups (higher ***CPP/FCSm*** values) perform better, as can be expected, but their size-dependence does not differ much from that of the lower performance groups.

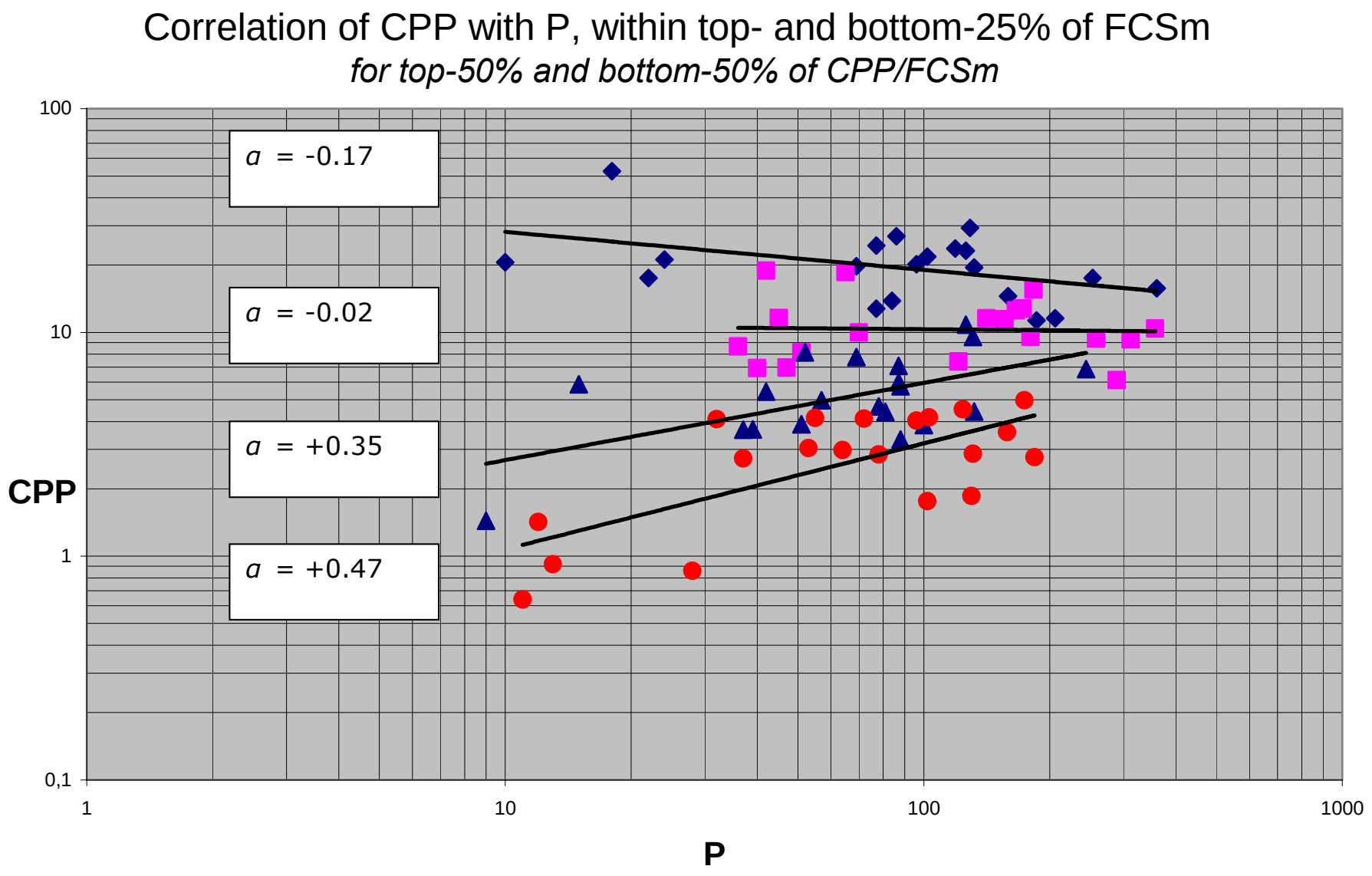

***Figure 5:*** *Correlation of citations-per-publication (**CPP**) with number of publications (**P**) for high field citation-density groups (top-25% of **FCSm**), divided in top-performance (top-50% of **CPP/FCSm**, diamonds) and lower performance (bottom-50% **of CPP/FCSm**, squares), and for low field citation-density groups (bottom-25% of **FCSm**), again divided in top-performance (top-50% of **CPP/FCSm**) (triangles) and lower performance (bottom-50% **of CPP/FCSm**, circles).*

We now try to explain these phenomena with our landscape model. Therefore, we investigate the behavior of the field citation-density itself as a function of size for both the high as well as the low field citation-density regions. The results are shown in Fig. 6 and reveal quite remarkable properties. For the *high* field citation-density groups a larger size does not significantly change (at most a very slight decrease) the average ***FCSm*** value. This means that for research groups operating in high field citation-density regions, a larger number of publications is realized within more or less the same high field citation-density regions. Fig. 7 shows that is the case for both high as well as lower performance groups. If we narrow down the top of the field citation-density, we find a slight decrease. This is understandable because a larger amount of publications will mostly imply extension toward regions with a somewhat lower field citation-density.

However, we also observe that for the groups in the *low* field citation-density regions a larger size implies a larger ***FCSm*** value. Thus, for groups operating in low field citation-density regions, a larger number of publications appears to go together with an 'expansion' into regions with higher field citation-density. Also here, the difference between high and low performance groups is not significant.

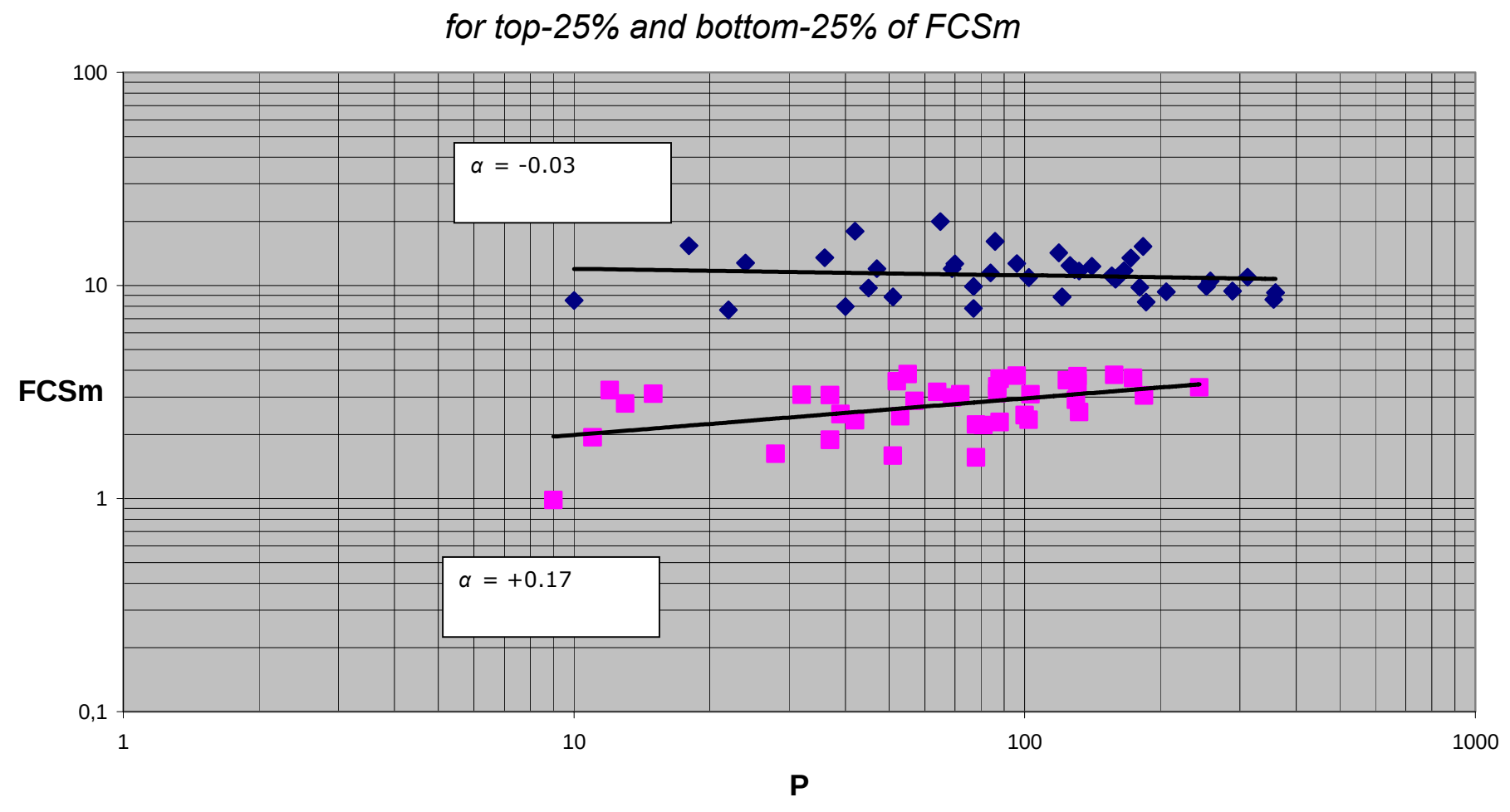

***Figure 6**:* Correlation of field citation density (***FCSm***) with size (***P***) for groups in fields with a high (diamonds) and a low (squares) citation density (***FCSm***)

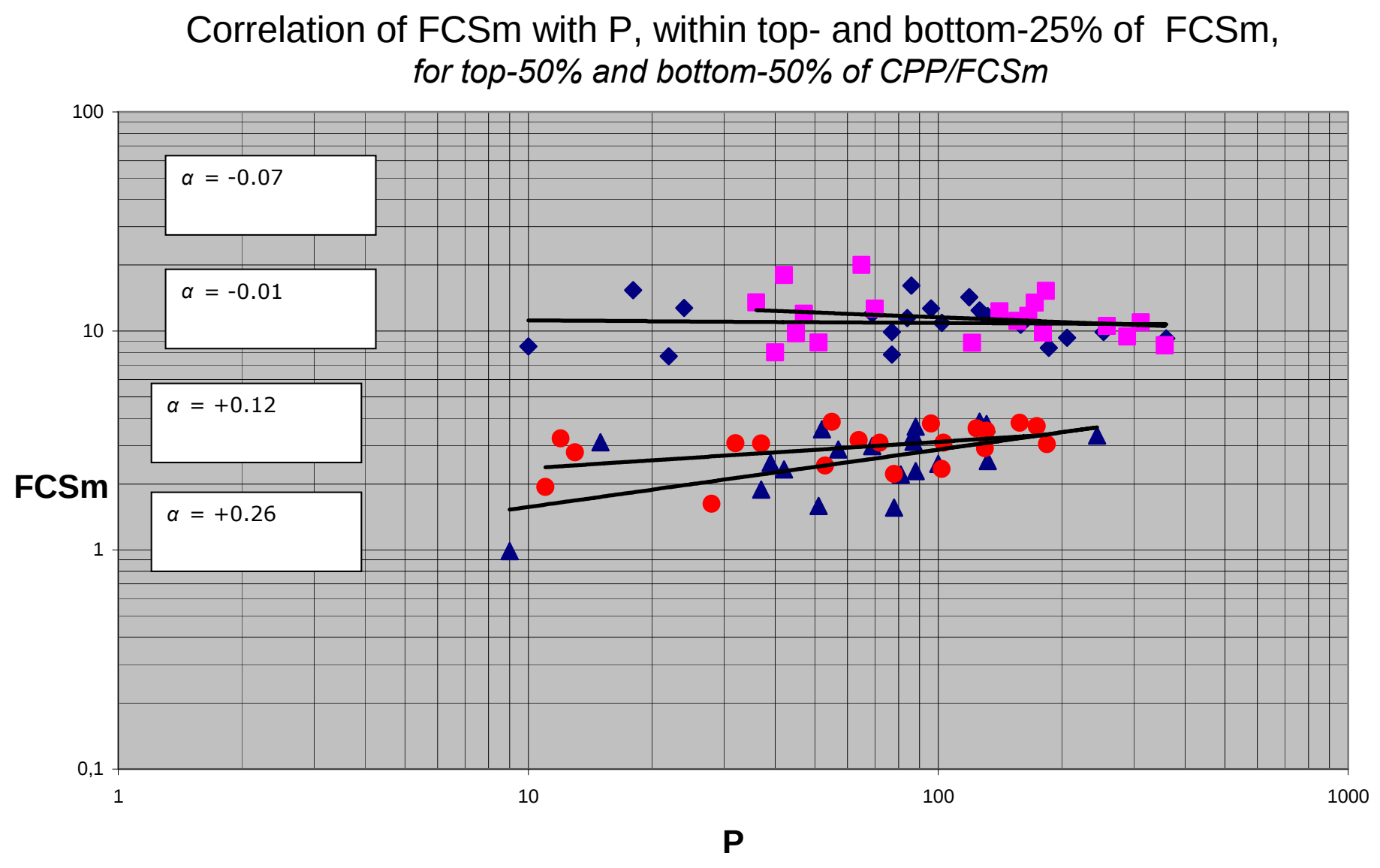

.

***Figure 7:*** *Correlation of field citation density (**FCSm**) with number of publications (**P**) for high field citation-density groups (top-25% of **FCSm**), divided in top-performance (top-50% of **CPP/FCSm**, diamonds) and lower performance (bottom-50% **of CPP/FCSm**, squares), and for low field citation density groups (bottom-25% of **FCSm**), again divided in top-performance (top-50% of **CPP/FCSm**) (triangles) and lower performance (bottom-50% **of CPP/FCSm**, circles).*

How does the average journal citation impact of a group relate to the field citation-density? The answer to the question is given by Fig. 8. We find that for the *high* field citation-density groups a larger size implies a *lower* average ***JCSm*** value. This implies for research groups operating in high field citation-density regions that a larger number of publications will lead to a somewhat lower average journal citation impact. This further completes the picture sketched by the previous observations: 'expanding in size' may take place with the same field citation-density region, but it will generally include publications in journals with a lower impact.  Fig. 9 shows that is particularly the case for the *high* performance groups. Clearly, lower performance groups publish generally in the lower impact journals and with a larger size they keep up publishing in these lower impact journals, so their average journal impact will not change significantly.

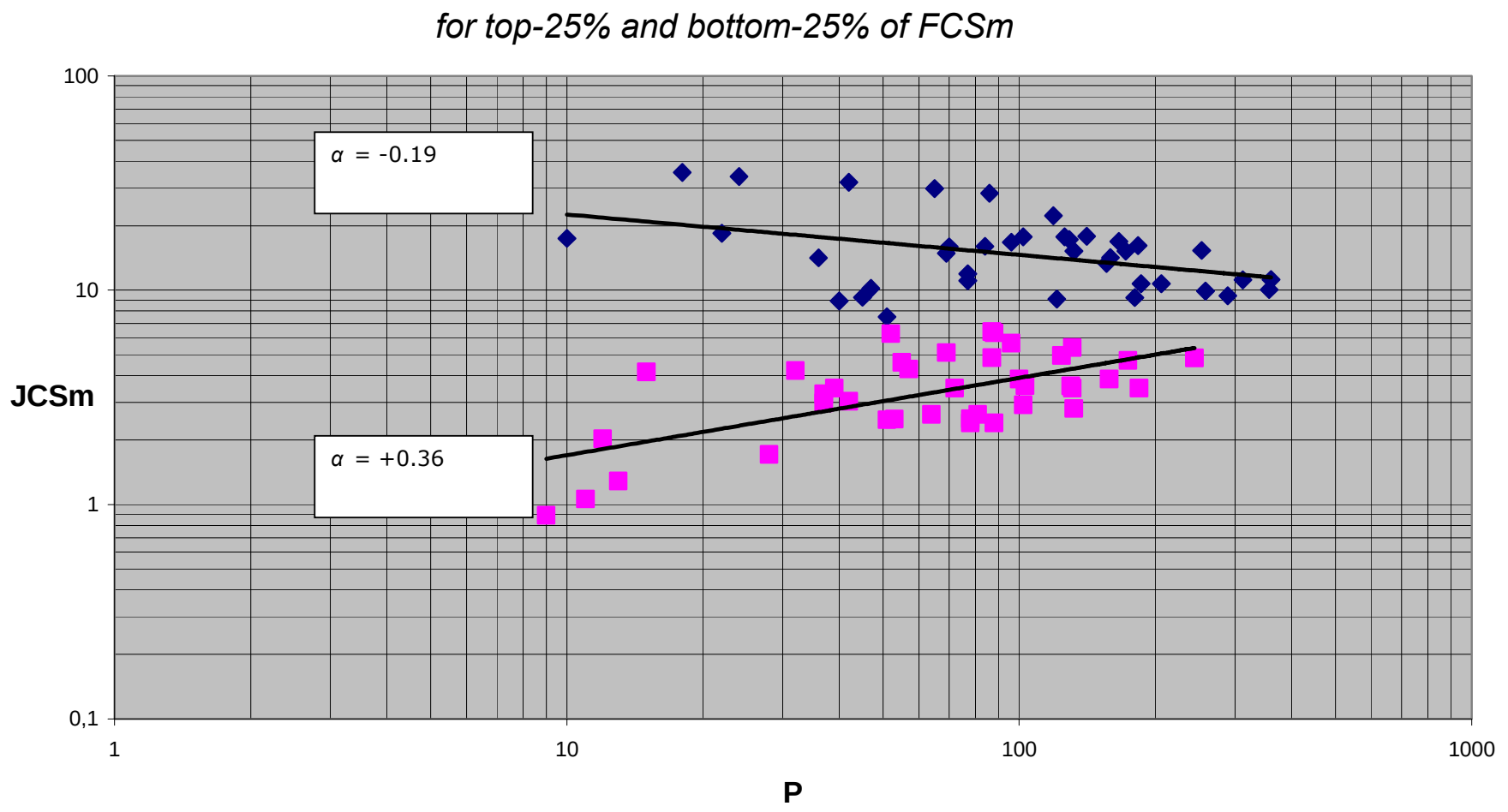

***Figure 8**: Correlation of journal impact (**JCSm**) with size (**P**) for groups in fields with a high (diamonds) and a low (squares) citation density (**FCSm**).*

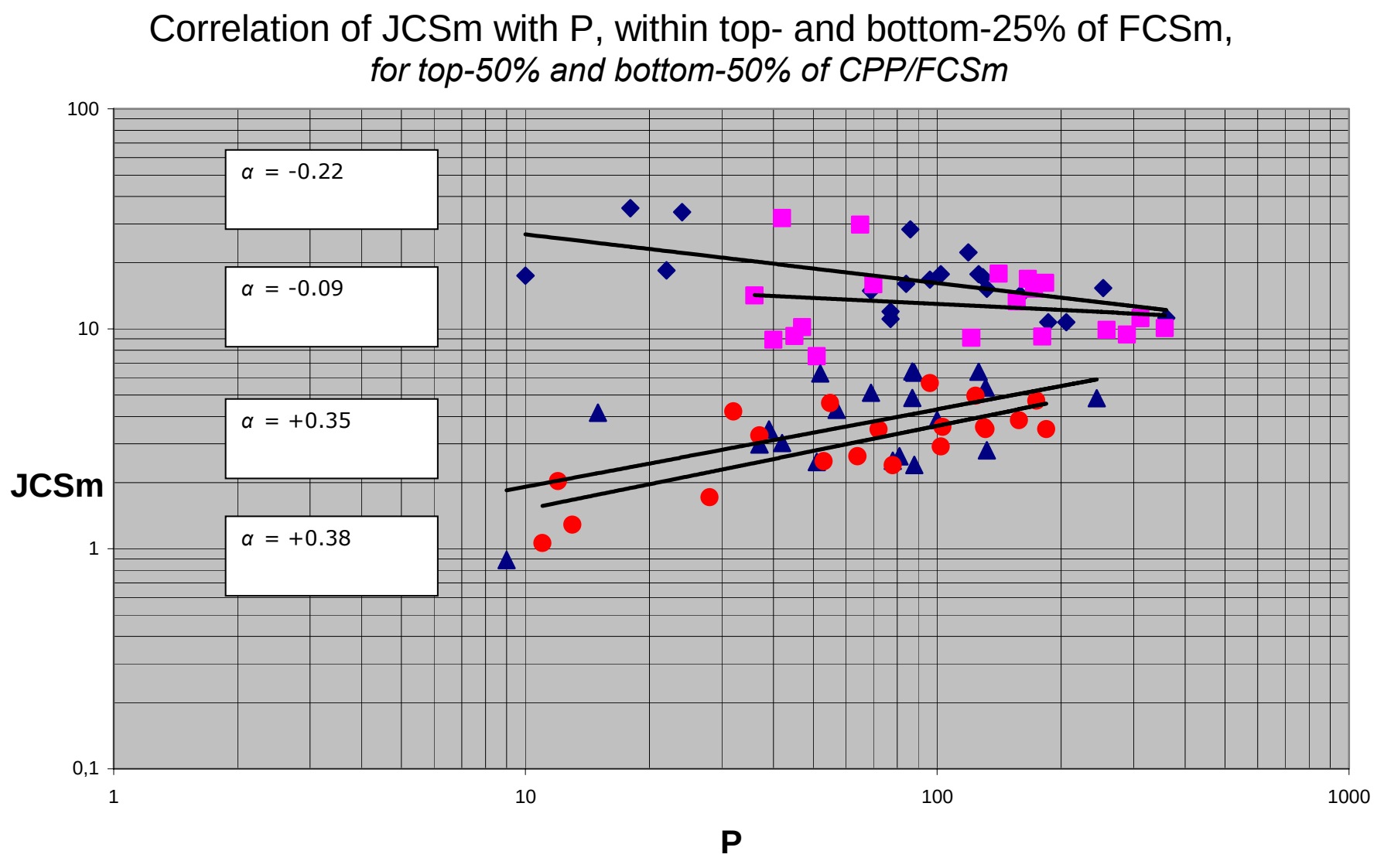

***Figure 9:** Correlation of field citation density (**JCSm**) with number of publications (**P**) for high field citation-density groups (top-25% of **FCSm**), divided in top-performance (top-50% of **CPP/FCSm**, diamonds) and lower performance (bottom-50% **of CPP/FCSm**, squares), and for low field citation density groups (bottom-25% of **FCSm**), again divided in top-performance (top-50% of **CPP/FCSm**) (triangles) and lower performance (bottom-50% **of CPP/FCSm**, circles).*

For groups in the *low* field citation-density regions however a larger size implies a considerably higher average ***JCSm*** value. Thus, for groups operating in low field citation-density regions a larger number of publications can be seen as an 'expansion' into regions with higher field citation-density as we saw earlier and at the same time as an expansion toward journals with a higher average impact. Again, in this behavior, the difference between high and low performance groups is not significant.

Next to the above dependence of the main bibliometric indicators on field citation-density, it is important to investigate specific interdependencies, particularly the interrelations between field citation-density and journal impact, and its influence on the total number of citations to a research group. We take the results presented in Fig. 3 and make a breakdown for both the high (top-25% of the ***FCSm*** distribution) as well as the low (bottom-25% of ***FCSm***) into the higher (top-50% of the ***JCSm*** distribution) and the lower (bottom-50% of ***JCSm***) journal impact, see Fig. 10.

Clearly we observe significant cumulative advantage (power law exponent α = 1.50) only for the groups with *low* field density and *low* average journal impact, there is no cumulative advantage for the high average journal impact fields. Obviously, high field citation density groups, whether they publish in lower or higher impact journals, receive considerably more citations than groups in the low field density regions. But they do not show significant cumulative advantage with size. As can be expected, for groups in both the high and low field citation densities, those groups that publish in higher impact journals will, for the group as a whole, be cited significantly more (see also our previous paper Van Raan 2006b, Fig.11).

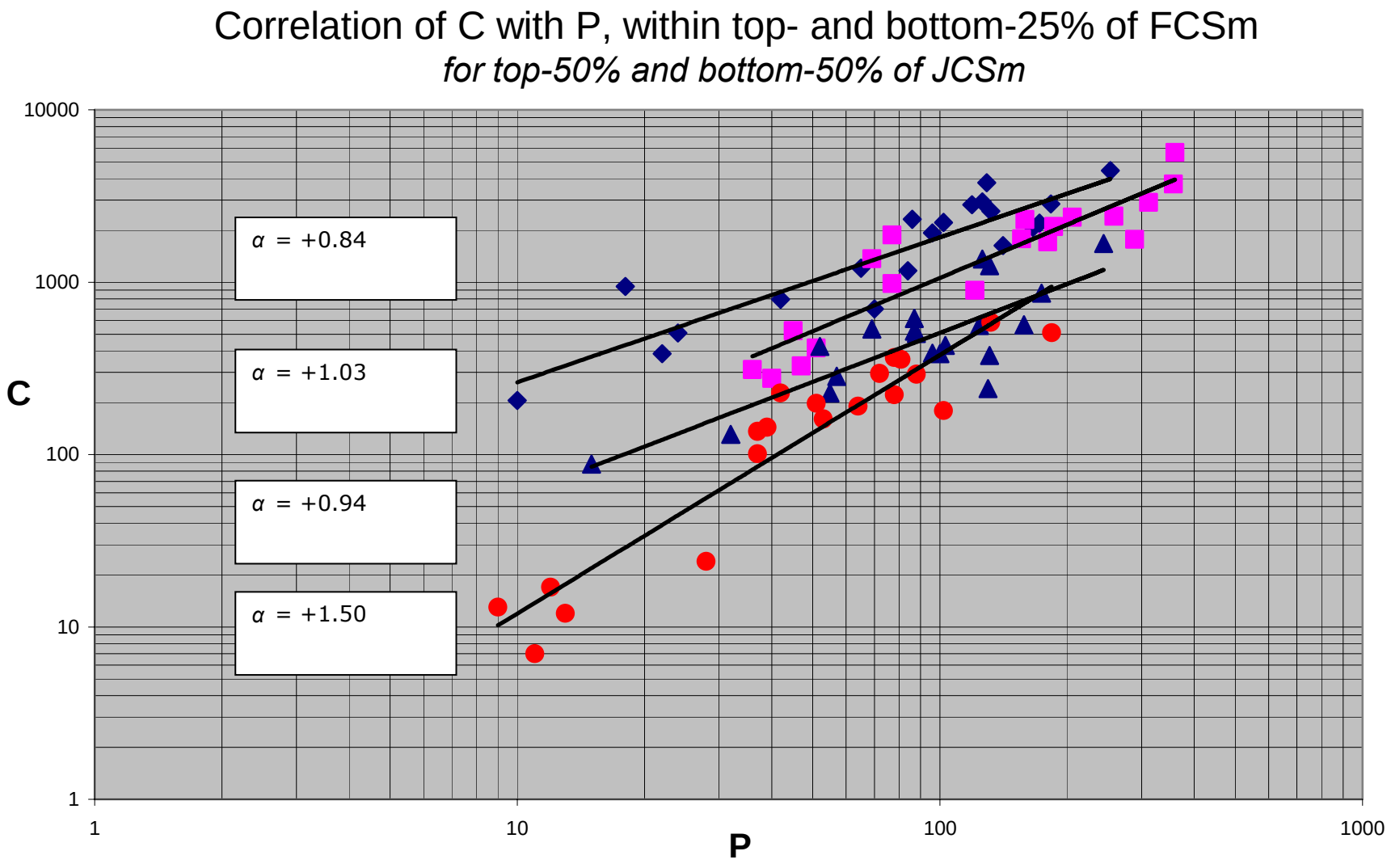

***Figure 10:*** *Correlation the number of citations (**C**) with the number of publications (**P**) for high field citation-density groups (top-25% of **FCSm**), divided in high journal impact (top-50% of **JCSm**, diamonds) and low journal impact (bottom-50% **JCSm**, squares), and for low field citation density groups (bottom-25% of **FCSm**), again divided in high journal impact (top-50% of J**CSm**) (triangles) and low journal impact (bottom-50% **JCSm**, circles).*

## Modelling the observations in the framework of the science landscape

We consider the field citation density as the main topological property of the bibliometric science system. We focus in our bibliometric approach of the science system on the most basic working units, the research group. These research groups have the following measurable attributes. First, the total number of publications of groups (***P***) is the measure for size of a group. Second, the total number of citations of groups (***C***) is the measure for the absolute impact of a group, it is the number of the citation linkages, or the amount of 'external wiring' of the group as a node in the network based on citations relations between groups. In network language, research groups are 'modules' of related publications. Thus, the indicator ***CPP*** represents the size-normalized wiring intensity of a group (module) in the network. Third, the research performance of a group measured by the field-normalized impact indicator ***CPP/FCSm*** represents the *fitness* of a group as a node in the *group*-network. It enables high performance groups to acquire substantially more links than lower performance groups, as illustrated by Fig. 5. In network terms there is a *preferential attachment* to high performance groups, i.e., other nodes, for instance newcomers, 'feel' the attractiveness (fitness) of a node and want to have a link with it.

Fourth, the field citation density ***FCSm*** can be seen as a *coarse-scale landscape property* of the groups in the networked science system. As discussed above, it is the main topological property of the bibliometric science system. For a research group, it is an *exogenous* property of the science system. Fifth, the journal citation density ***JCSm*** can be seen as a fine-tuning within this coarse scale structure. Compared to the field citation density, the journal citation density is a more *endogenous*, i.e., group-specific attribute, a 'basic facility' used by the group to put itself in a better position. By analogy with a social context one could think of ***FCSm*** as related to the social layer or social environment in which a family lives, and ***JCSm*** as the education level. A higher level of education (***JCSm***) increases -but does not guarantee- the probability to improve the family status within the social environment (***FCSm***) in order to reach a higher income (***CPP***). But this is not an automatism, and with a relatively low education level one has still has a chance for a high income.

In this paper we show that total number of citations received by research groups increases in a cumulatively advantageous way as a function of size only for groups publishing in fields of low citation density, regardless of performance. In our previous study (van Raan 2006b) we found that, also regardless of performance, larger groups have fewer not-cited publications. By distinguishing again between top- and lower-performance groups, we found that particularly for the lower performance groups the fraction of not-cited publications decreases considerably with size. We presented in this previous paper a model with two independent mechanisms to explain the observed phenomena. In mechanism A, *self*-citations 'promote' *external* citations in a non-linear relation with size, thus decreasing the number of not-cited publications. Mechanism B concerns the role of the field citation density. The previous paper dealt with mechanism A, in this paper we focus on mechanism B, the influence of the field citation density.

The working of mechanism B can be explained as follows, see Fig. 11. We take as an example a research group in a field of low citation density. Our empirical results presented in Fig. 6 show that the average field citation density of a group slowly increases with size (number of publications), but particularly on a more finer scale the *journal citation density* ***JCSm*** will increase more strongly, see Fig. 8. Thus, for groups in *low* citation density fields, a larger number of publications implies a higher probability of expansion into higher citation densities. This means that the relative increase of citations $(\Delta C)/C$ will be larger than the relative increase of publications $(\Delta P)/P$ which can be calculated as follows. From the empirical results presented in Fig. 8 we deduce:

$$(JCSm) = kP^{\beta} \qquad \text{(Eq. 1)}$$

with $\beta \sim + 0.4$ for groups with a *low* field citation density and $\beta \sim - 0.2$ for groups with a *high* field citation density. Given the significant correlation between ***JCSm*** and ***CPP*** for the entire set of 157 chemistry research groups, $(CPP) = 1.13\ (JCSm)^{\gamma}$ with $\gamma = 0{,}97$ (van Raan 2006b, Fig. 9a), and reminding that the indicator-symbol ***CPP*** stands for citation per publication which is $C/P$, we may replace in good approximation $(JCSm)$ by $C/P$. By differentiating Eq. 1 we find that for the *low* field citation-density groups a relative increase of publications is related to a relative increase of citations as follows:

$$\Delta P/P = \alpha^{-1}\ (\Delta C)/C, \text{ with } \alpha = \beta+1 \sim +1.4 \qquad \text{(Eq. 2)}$$

and after elementary algebra we find a power-law dependence of the number of citations received by groups ($C$) with size $P$, with 'cumulative advantage' (power law exponent $\alpha$ *larger* than 1)

$$C(P) = aP^{\alpha}\ ,\ \alpha \sim +1.4 \qquad \text{(Eq. 3)}$$

which corresponds well with our findings presented in Fig. 3.

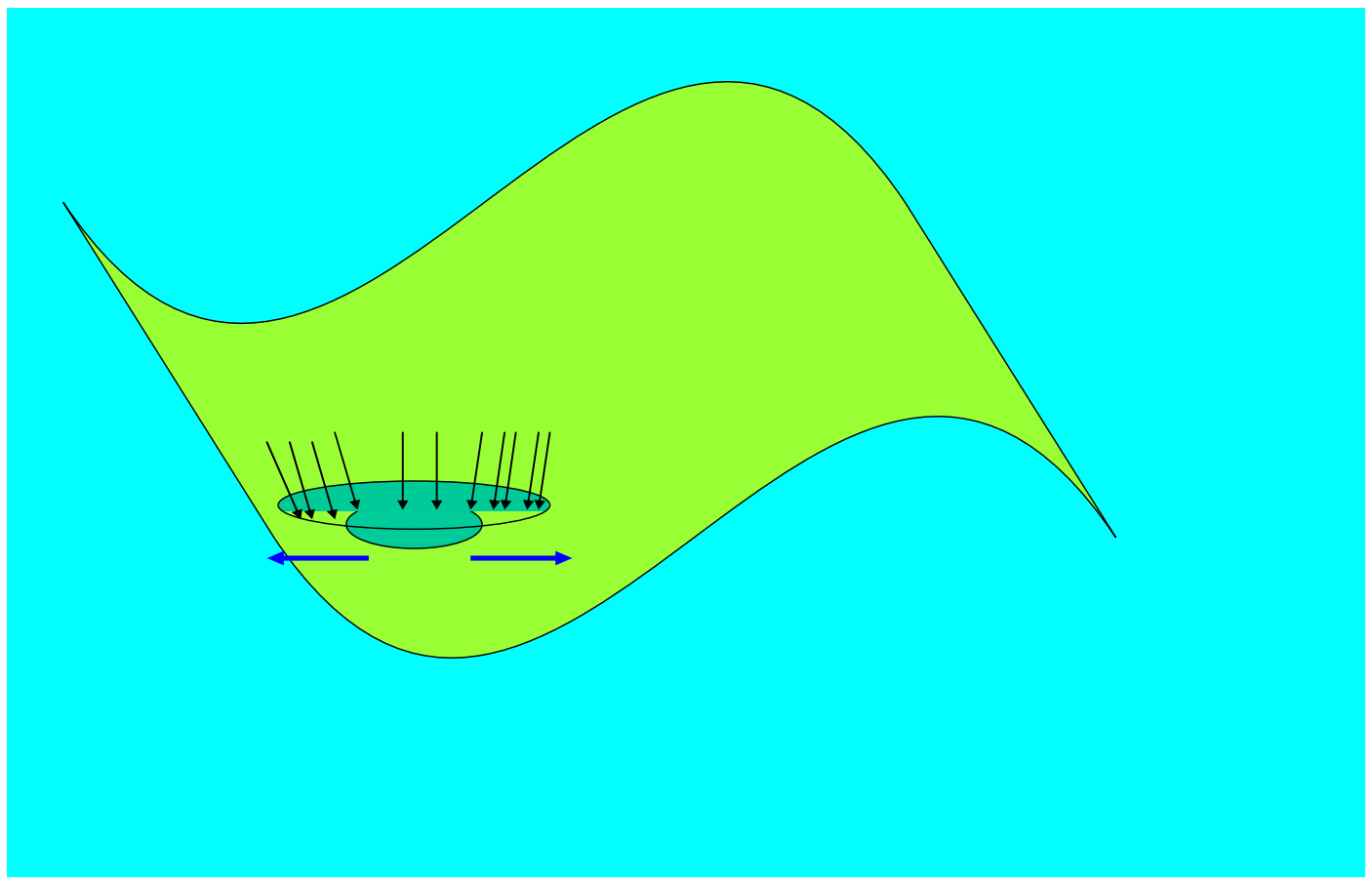

***Figure 11***: *Larger size increases the probability of extension toward regions of higher citation density.*

For groups in *high* citation density fields, a larger number of publications will not significantly change the larger scale average field citation density (see Fig. 6) but the finer scale journal citation density will decrease, be it not very strong. Thus, in the case of groups in *high* citation density fields the relative increase of citations $(\Delta C)/C$ will be somewhat smaller than the relative increase of publications $(\Delta P)/P$.

With a similar mathematical procedure as above we find for the *high* field citation-density groups again the same power law relation but now without cumulative advantage (power law exponent ***α*** *smaller* than 1):

$\boldsymbol{C}(\boldsymbol{P}) = a\boldsymbol{P}^{\boldsymbol{\alpha}}$, $\boldsymbol{\alpha} \sim +0.8$

which again corresponds well with our findings presented in Fig. 3.

In this paper we showed that the science system modelled as a landscape described by the concept of field citation-density, reveals despite 'microscopic randomness' (e.g., the probability that an individual publication will be cited) a few scaling rules that can explain several important features of this complex system, especially the size-dependence of several main bibliometric indicators for a large set of research groups while distinguishing between top-performance and lower performance groups. The basic scaling rule is the relation between the field citation density (***FCSm***) and size in terms of number of publications (***P***), from which the relation between the absolute number of citations (***C***) and size can be deduced using a model in which the working of the field citation density in the science landscape is explained.

***Acknowledgements***

The author would like to thank his CWTS colleague Thed van Leeuwen for the data collection, data analysis and calculation of the bibliometric indicators for the set of research groups.